\documentclass[12pt,preprint]{emulateapj}

\shorttitle{Galactic Anticenter Stellar Structure Spectroscopy}
\shortauthors{Crane et al.}

\begin{document}

\title{Exploring Halo Substructure with Giant Stars:\\
Spectroscopy of Stars in the Galactic Anticenter Stellar Structure}

\author{Jeffrey D. Crane\altaffilmark{1}, Steven R. Majewski\altaffilmark{1}, 
Helio J. Rocha-Pinto, Peter M. Frinchaboy, Michael F. Skrutskie, and 
David R. Law}
\affil{Department of Astronomy, University of Virginia,
    Charlottesville, VA 22903}
\email{jdc2k, srm4n, hjr8q, pmf8b, mfs4n, drl5n@virginia.edu}

\altaffiltext{1}{Visiting Astronomer, Kitt Peak National Observatory,
National Optical Astronomy Observatory, which is operated by the Association
of Universities for Research in Astronomy, Inc. (AURA) under cooperative
agreement with the National Science Foundation.}


\begin{abstract}

To determine the nature of the recently discovered, ring-like stellar 
structure at the Galactic anticenter, we have collected spectra of a set of 
presumed constituent M giants selected from the 2MASS point source catalog. 
Radial velocities have been obtained for stars spanning $\sim100\degr$, 
exhibiting a trend in velocity with Galactic longitude and an estimated 
dispersion of $\sigma_{v}=20 \pm 4$ km s$^{-1}$. A mean metallicity 
[Fe/H]$=-0.4 \pm 0.3$ measured for these stars combines with previous
evidence from the literature to suggest a population with
a significant metallicity spread. In addition, a curious alignment
of at least four globular clusters of lower mean metallicity is noted to 
be spatially and kinematically consistent with this stellar distribution.
We interpret the M giant sample position and velocity variation with 
Galactic longitude as suggestive of a satellite galaxy currently undergoing 
tidal disruption in a non-circular, prograde orbit about the Milky Way. 

\end{abstract}



\section{Introduction}

An extensive, coherent stellar structure has revealed itself from behind 
the obscuring veil of the disk in the direction of the Galactic anticenter. 
First identified as an overdensity of blue stars in the Sloan Digital Sky 
Survey \citep{newberg} (hereafter N02), this structure was subsequently 
studied by \citet{ibata} (hereafter I03) using INT WFS photometry, and by 
\citet{yanny} (hereafter Y03) using SDSS photometry and spectroscopy. 
\citet{majewski} (hereafter M03) found evidence for the structure in their 
study of M giants selected from 2MASS. \citet{helio} (hereafter R03) also 
made use of 2MASS M giants in a targeted search for evidence of the 
structure in numerous 4$\degr$ fields in the second and third Galactic 
quadrants. Following Y03, we refer to this Galactic anticenter stellar 
feature as the ``Mon'' structure, although future work may prove this to be 
an inappropriate name (if a dwarf galaxy, the core does not seem to be in 
Monoceros).

Between I03 and Y03, evidence for the Mon structure has been found in fields
spanning $-27\degr<b<+30\degr$ and $122\degr<l<225\degr$ with possible 
extension to $l=90\degr$. N02 estimate the distance of their initial 
detection at $\sim$11 kpc from the Sun, with a Galactocentric distance of 
$\sim$18 kpc. I03 determined that the structure's 
Galactocentric distance changes from 18 kpc at $l=221\degr$ to 14 kpc at
$l=149\degr$. R03 have been able to trace the structure clearly over 
$150\degr<l<220\degr$ using 2MASS M giants, with less obvious, but still
strong evidence for coherence over $100\degr<l<270\degr$, and distances
consistent with those previously reported. Scaleheights for the structure 
have been estimated at $<3$ kpc, $0.75\pm0.04$ kpc, and $1.3\pm0.4$ kpc, 
but variable with Galactic longitude (Y03, I03 and R03, respectively). 
The radial thickness has been estimated to be $<4$ kpc (Y03) and 
$\sim2$ kpc (I03).

Based on velocities from both targeted and serendipitously collected SDSS
spectra, and on the assumption of a circular orbit, Y03 report a $110\pm25$
km s$^{-1}$ prograde rotation for the Mon system, and velocity dispersions 
too narrow (22--30 km s$^{-1}$) to be identified with any known Galactic 
component. They estimate a preliminary metallicity of [Fe/H]$=-1.6\pm0.3$ 
for their blue stars, based on main sequence turnoff colors and 
\ion{Ca}{2} (K) line strengths from spectra with $S/N\sim10$ at $R\sim2000$.

Despite this flurry of study over the last year, the nature of the Mon
structure remains unsettled. I03 suggest a number of scenarios including 
a tidally disrupted satellite galaxy, an outer spiral arm, or even a 
resonance induced by an asymmetric Galactic component. However, I03 hold
slight preference for the interpretation that the Mon structure is a 
perturbation of the Galactic disk resulting from old, repeated warpings
due to the influence of the Magellanic clouds and/or the Sagittarius (Sgr)
dwarf. Y03 prefer the interpretation that the structure is a tidally 
disrupted satellite. \citet{helmi} have examined numerical simulations of 
tidally disrupted satellite galaxies in roughly coplanar orbits with disk 
galaxies. Although not specifically targeted to explain the Mon 
structure, these simulations provide interesting insight into the 
structure of tidally disrupted objects at different evolutionary epochs. 
Specifically, young tidal arcs are azimuthally limited and exhibit 
a strong radial velocity gradient, while older tidal arcs mix to form 
shells that show no azimuthal velocity trend.

In this work, the nature of the Mon structure is explored spectroscopically 
over a wide area.


\section{Target Selection}

\begin{figure}
\epsscale{1.00}
\plotone{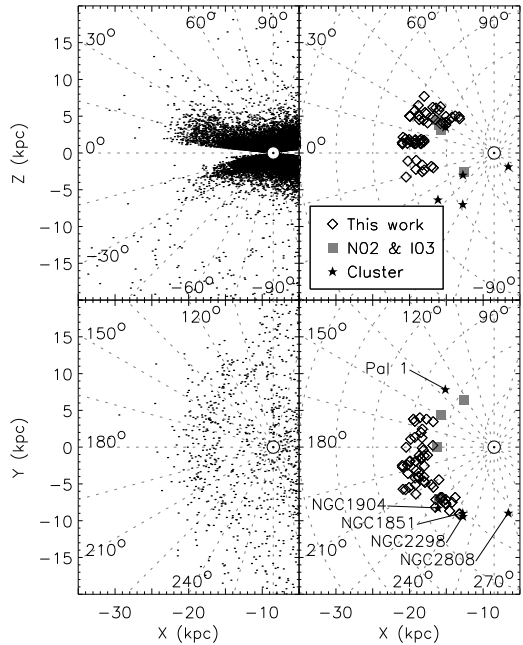}
\caption{Distributions of M giants shown in Galactocentric Cartesian
coordinates, where the Sun is at $(X,Y,Z)=(-8.5,0,0)$ kpc. 
Galactic longitude and latitude lines are shown, as well as 
curves of constant Galactocentric radius. 
Illustrative slices through the 2MASS M giant distribution are shown on the 
left: $-5<Y<+5$ kpc for the upper panel and $+4<Z<+7$ kpc for the lower 
panel. Note the asymmetric $X-Z$ distribution, where more stars appear 
to reside above than below the plane at $X<-15$ kpc (this is at least partly 
due to increased reddening in the Southern hemisphere; see 
R03). The Mon feature is also identifiable in the $X-Y$ 
projection. Positions of stars observed spectroscopically are shown on 
the right, as are several globular clusters of interest.
\label{xyzplot}}
\end{figure}

Targets were selected based on the detection of the Mon structure in M03's
work to trace the tidal tails of the Sgr dwarf 
galaxy. In their Figure 10, the Mon structure is apparent as an asymmetric 
concentration of stars near the Galactic plane at 
$\Lambda_{\sun}\sim165\degr$ in the Sgr plane coordinate system. At the 
longitudes probed by that Figure, it is apparent that the Mon structure is
more concentrated to the northern Galactic hemisphere, an observation 
noted by previous authors and also suggested by the distribution of 2MASS 
stars shown in Figure~\ref{xyzplot}.

The sample consists of M giants selected using the two-color discrimination 
technique described in M03. Stars in significantly reddened directions 
($0.555<E[B-V]$ and $-5\degr<b<+5\degr$) were excluded to minimize 
photometric parallax errors. In addition, we imposed the restrictions 
$(J-K_{s})>1.0$ and $d_{\sun}>10$ kpc (with distances estimated from the 
Sgr core absolute magnitude-color relation from M03). The former criterion 
removes low-metallicity Milky Way giants (and yields the brightest targets), 
and the latter removes the bulk of the disk stars and focuses on
Galactocentric radii reported to be occupied by the Mon structure.
Selection of stars in the noted concentration around 
$\Lambda_{\sun}\sim165\degr$ in M03's Figure 10 practically limits the
sample's Galactic longitude range since that plot represents only a 14 kpc 
thick slice through the Mon feature. 

The distribution of targets is shown in Figure~\ref{xyzplot}.
We concentrate primarily on the Northern Galactic 
hemisphere, where the largest overdensity appears in the smoothed 2MASS 
M giant distribution. The sample's radial coverage is larger in the 
direction toward $l\sim200\degr$, where R03 also note increased structure 
depth.


\section{Data Collection and Reduction}

Data were collected UT 2003 April 17--25 using the GoldCam Spectrograph on 
the 2.1-meter telescope at Kitt Peak National Observatory. The spectra 
cover $7800<\lambda<9000$\AA~ with instrumental $R\sim3300$.

Following overscan correction, trimming, and bias subtraction, the data were
flatfielded using quartz lamp exposures that had been spectrally normalized
to remove the large-scale spectral response. This latter step adequately
removed fringing. Spectra were extracted and wavelength-calibrated 
using comparison lamp spectra taken at the same telescope position 
as the targets.

Using our own software, radial velocities were measured by cross-correlating 
the fourier-filtered spectra against a variety of different standards of 
late spectral type. Average velocity uncertainties of $\pm2.7$ km s$^{-1}$ 
were achieved for target spectra with mean peak $S/N\sim41$ correlated 
against 26 standard star observations. 


\section{Velocities}

\begin{figure*}
\epsscale{1.00}
\plotone{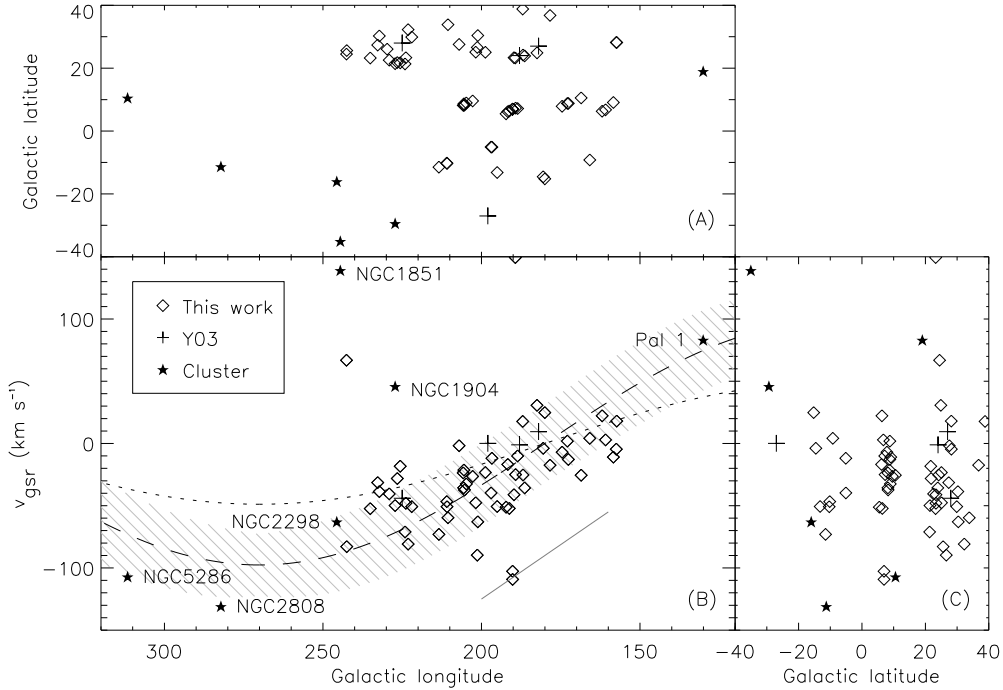}
\caption{The distributions of targets in position (Panel A) and 
Galactocentric radial velocity. Observed velocities are 
corrected to Galactic standard of rest ($v_{gsr}$) by assuming a solar 
apex of $(\alpha,\delta)=(18h,30\degr)$ at 20.0 km s$^{-1}$ and an LSR 
rotation of 220 km s$^{-1}$. The expected velocity dispersion for the thick 
disk is represented by the region filled with diagonal lines. 
The dotted line shows the $v_{gsr}$ corresponding to Y03's reported
$110\pm25$ km s$^{-1}$ circular velocity. The dashed curve represents 
the expected $v_{gsr}$ for an object orbiting circularly at 
r$_{GC}=18$ kpc with $v_{circ}=220$ km s$^{-1}$, which is a better match 
to the data. The sold line shows the trend of the \ion{H}{1} feature 
identified by \citet{simonson}. Four globular clusters appear to be 
positionally and kinematically similar to the Mon structure, while 
NGC~1851 and NGC~1904 are positionally interesting, but well off 
the Mon M giant velocity trend. Velocity uncertainties for the 58 M giants 
are $\pm2.7$ km s$^{-1}$.
\label{vgsr}}
\end{figure*}

The Mon M giant sample shows (Figure~\ref{vgsr}) a positive velocity 
gradient with decreasing Galactic longitude, consistent with prograde 
rotation. While this agrees qualitatively with the results of Y03, at 
similar longitudes the M giant radial velocities are systematically lower 
by $\sim20$ km s$^{-1}$. We have found (S. R. Majewski et al., in 
preparation) a similar offset with 
Y03 observations of the Sgr tidal tails; this problem is discussed further 
in that reference. A third-order polynomial fit to the velocity trend shows 
that the locus passes through $v_{gsr}\sim-16$ km s$^{-1}$ at $l=180\degr$, 
which indicates a slightly non-circular orbit.  The velocity locus passes 
through $v_{gsr}=0$ at $l=166\degr$. The present latitude coverage is less 
uniform (reflecting only our target selection), but is consistent with no 
velocity gradient in that dimension.

Of the main Galactic components, the thick disk should be the dominant
contributor of M giants in the regions represented by most of our stars, 
with the thin disk contributing significantly less and the halo 
contributing negligibly \citep{siegel}. Using
$\langle\sigma_{U},\sigma_{V},\sigma_{W}\rangle=\langle38,25,20\rangle$ 
and $\langle60,45,40\rangle$ km s$^{-1}$ for the thin and thick disks 
\citep{gilmore}, we estimate the expected line-of-sight velocity 
dispersions for these components at $R_{gc}=18$ kpc toward $l=200\degr$ 
to be $\sim$40 and $\sim$64 km s$^{-1}$, respectively. In contrast, a 
line-of-sight velocity dispersion of $\sigma_{v}\sim20 \pm 4$ km s$^{-1}$ 
for 53 stars in the Mon structure was determined about the polynomial fit 
after using a $2.5\sigma$ rejection threshold that removed 5 stars. 
This is substantially smaller than expectation for the Galactic components.

It is interesting to note that structure \citep{tamanaha} in the 21-cm 
\ion{H}{1} emission near the Galactic anticenter has previously been 
suggested \citep{simonson} to imply the presence of a tidally disrupted 
satellite galaxy, although subsequent work \citep{bignami,burton} did not 
favor this interpretation. A connection to the Mon--Milky Way
interaction is suggested by the similar spatial distributions, but the
shifted velocity trend (Figure~\ref{vgsr}) implies that establishing this
will require consideration of more complicated interactions with 
the Milky Way's gaseous disk.

\section{Metallicity}

\begin{figure}
\epsscale{1.00}
\plotone{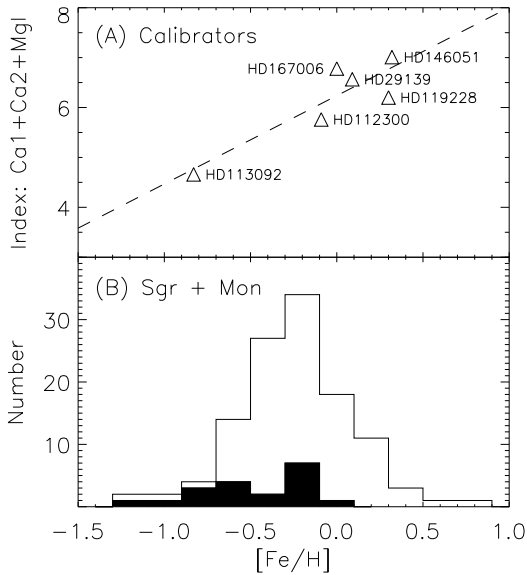}
\caption{Panel A shows the linear fit to our metallicity calibrators using
a direct summation of the \citet{diaz} indices. Panel B shows the result
of applying the fit to 117 high $S/N$ Sgr (unfilled histogram) and 19 Mon
(filled histogram) M giants.
\label{metal}}
\end{figure}

As R03 point out, the mere presence of M giants in the Mon structure provides 
circumstantial evidence for a relatively metal-rich component. We attempted 
to estimate the metallicities of our sample stars using the \ion{Ca}{2} 
index CaT* defined by \citet{cenarro1}. However, their CaT* fitting functions 
\citep{cenarro2} are not metallicity sensitive for stars in the 
temperature range of our sample. Attempts to apply the principal component 
analysis technique of \citet{diaz} met with similarly ambiguous 
results since that work was not calibrated to sufficently late spectral 
types. In the end, a much-simplified method was adopted to estimate the 
metallicities of a representative subsample of stars. A direct summation 
of \ion{Ca}{2} $(\lambda\lambda 8498, 8542$\AA) and \ion{Mg}{1} 
$(\lambda\lambda 8807$\AA) indices \citep{diaz} was calculated for a set
of bright, red calibrator stars. A linear fit was then derived between 
the combined indices of these stars and their published metallicities 
(Figure~\ref{metal}). Calibrators were restricted to $0.90<(J-K_{s})<1.1$,
log$(g)<2.5$, and $-0.83<$ [Fe/H] $<+0.32$. Applying the linear 
fit for these data to Mon stars in the same color range and with 
$S/N>40$ yields $\langle$[Fe/H]$\rangle =-0.4\pm0.3$. As a relative 
reference, the 
metallicity determined for a larger number of Sgr M giants observed during 
the same run is found to be [Fe/H] $=-0.2\pm0.3$, slightly more metal-rich 
(but still within the uncertainty) than the [Fe/H] $=-0.4$ reported by 
\citet{layden} for the most recent episode of Sgr star formation. 
Regardless of absolute calibration uncertainties, the Mon M giants appear 
to be of comparable, but slightly lower metallicity than Sgr M giants.

\section{Discussion}
\label{discussion}

We favor the interpretation that the Mon structure is a satellite galaxy 
currently undergoing tidal disruption and similar to the Sgr system for the
following reasons:

(1) The velocity trend with longitude indicates that the structure is in a 
non-circular orbit, but with low eccentricity. \citet{bosch} find a preference 
for higher eccentricities in distributions of satellites born in a 
spherical potential, and recent theoretical work \citep{colpi,hashimoto}, 
indicates that circularization of these orbits by dynamical friction is 
probably insignificant. However, \citet{taffoni} demonstrate how 
sensitively the tidal disruption of satellites depends on their orbital 
eccentricity, so that those currently remaining intact or undergoing 
disruption now should also be those originally born on near-circular 
orbits. Thus, while I03 believe the near-circular orbit casts doubt on the 
disrupted satellite scenario, it does not rule it out. Indeed, increased
circularity might be expected for objects accreted at later epochs.

(2) We estimate the velocity dispersion of our sample to be 
$\sigma_{v}=20 \pm 4$ km s$^{-1}$, notably lower than Y03's 27, 22, 30, 
and 30 km s$^{-1}$ for their four fields. Nevertheless, in all cases the 
dispersion is lower than would be expected of any known Galactic component. 

(3) We estimate [Fe/H] $=-0.4\pm0.3$ for our sample, showing the existence of 
a relatively enriched population. Considering the [Fe/H] $=-1.6\pm0.3$ 
reported by Y03, a spread in metallicity is implied, and suggestive of 
multiple epochs of star formation, as seen in the Sgr dwarf \citep{layden}. 
While our estimate is also consistent with the disk metallicity 
([Fe/H]$\sim-0.6$ at $R_{gc}\sim18$ kpc) expected from extrapolation of the 
open cluster metallicity gradient \citep{chen}, much of our sample is so 
high above the Galactic plane that we find it difficult to understand how 
all of the M giants could be associated with the disk, unless a very large 
warp is invoked. The Galactic \ion{H}{1} warp \citep{nakanishi} is not large 
enough, nor is it in the same orientation, to explain our sample as a disk 
feature unless the stars in the warp do not trace the gas.

I03 suggest that a tidal stream should fill the radial space between 
perigalacticon and apogalacticon after a sufficiently long period of time. 
\citet{helmi} note that tidal arcs become tenuous and difficult to recognize
relatively quickly after their first pericentric passage; older ring-like
features eventually form shells around the central galaxy and therefore 
would show no velocity trend with azimuthal angle. Thus, the velocity
coherence, velocity trend with longitude, thin radial thickness (estimated
by Y03 and I03), and separation from the edge of the disk (Figure 4 in R03)
supports not only the hypothesis that the Mon structure is a tidal stream, 
but also one that is relatively dynamically young. 

(4) At least four globular clusters are noted (Figure~\ref{vgsr}) that may 
be associated with the stream based on \textit{both} position and 
kinematical data \citep{harris}. This string-like orientation of low
latitude clusters is unusual in the outer Milky Way Galactic Cluster
system, but resembles the correlation of clusters to the tidal arms of the
Sgr system \citep{bellazzini}. The correlated cluster metallicities 
exhibit a spread that is also similar to that of the Sgr clusters --- 
[Fe/H]~$=-0.60$, $-1.15$, $-1.67$, and $-1.85$ for \objectname[Pal]{Pal~1},
\objectname[NGC]{NGC~2808}, \objectname[NGC]{NGC~5286}, and
\objectname[NGC]{NGC~2298}, respectively. We note an additional cluster,
\objectname[BH]{BH~176}, that is suitably placed in position, but for which 
no velocity or metallicity information is currently available. If these
clusters are genuinely associated with the Mon structure, then we would 
conclude with even greater certainty that it is in fact a Sagittarius-like 
dwarf galaxy undergoing tidal disruption by the Milky Way.


\acknowledgments

We thank Butler Burton for an insightful correspondence, and
Matt Garvin, Kathryn Johnston, Megan Kohring, Christopher Palma,
and Michael Siegel for software that aided our analyses. This
publication makes use of data products from the Two Micron All Sky Survey,
which is a joint project of the University of Massachusetts and the Infrared
Processing and Analysis Center/California Institute of Technology, funded by
the National Aeronautics and Space Administration and the National Science
Foundation.


\end{document}